\documentclass[aps,twocolumn,pra,floatfix,superscriptaddress,10pt]{revtex4-1}
\usepackage{times}
\usepackage{amsmath,amssymb}
\usepackage{graphicx}
\graphicspath{{graphics/}}
\usepackage{braket}
\usepackage{float}
\usepackage{tikz}
\usetikzlibrary{positioning}

\usepackage{hyperref}

\DeclareMathOperator{\Tr}{Tr}
\newcommand{\bok}[3]{\langle #1 | #2 | #3 \rangle}

\begin{document}

\begin{abstract}
We present a numerical method to approximate the long-time asymptotic solution $\rho_\infty(t)$ to the Lindblad master equation for an open quantum system under the influence of an external drive. The proposed scheme uses perturbation theory to rank individual drive terms according to their dynamical relevance, and adaptively determines an effective Hamiltonian. In the constructed rotating frame, $\rho_\infty$ is approximated by a time-independent, nonequilibrium steady-state. This steady-state can be computed with much better numerical efficiency than asymptotic long-time evolution of the system in the lab frame. We illustrate the use of this method by simulating recent transmission measurements of the heavy-fluxonium device, for which ordinary time-dependent simulations are severely challenging due to the presence of metastable states with lifetimes of the order of milliseconds. 
\end{abstract}

\title{Adaptive Rotating-Wave Approximation for Driven Open Quantum Systems}
\author{Brian Baker}
\affiliation{Department of Physics and Astronomy, Northwestern University, Evanston, Illinois 60208, USA}

\author{Andy C.~Y.~Li}
\affiliation{Department of Physics and Astronomy, Northwestern University, Evanston, Illinois 60208, USA}

\author{Nicholas Irons}
\affiliation{Department of Physics and Astronomy, Northwestern University, Evanston, Illinois 60208, USA}

\author{Nathan Earnest}
\affiliation{The James Franck Institute and Department of Physics, University of Chicago, Chicago, Illinois 60637, USA}

\author{Jens Koch}
\affiliation{Department of Physics and Astronomy, Northwestern University, Evanston, Illinois 60208, USA}

\maketitle

\section{Introduction\label{Introduction}} 

Recent advances in the design of quantum systems such as superconducting qubits \cite{Blais2004,Martinis2005,koch07,Clarke2008,Girvin2009,Brooks2013,Devoret2013}, trapped ions \cite{Cirac1995,Haffner2008,Monroe2013}, and optical lattices \cite{Brennen1999,Bloch2008} have intensified the spotlight on the goal of realizing a quantum computer. Essential to this goal is the capability to control quantum systems coherently, while minimizing the influence of noise. Qubit control via an external drive has been extensively studied both theoretically and experimentally, particularly for gate operations \cite{Chow2010,Barends2014}, initialization \cite{Geerlings2013}, and readout \cite{Hofheinz2008, Vijay2012}. 

Predictions of the nonequilibrium dynamics of driven open quantum systems can often be based on framework of the Lindblad master equation \cite{Gorini1976,Lindblad1976,Francesco07}. In most cases, solving this equation has to rely on numerical methods and faces multiple challenges, including Hilbert-space size and the resulting memory requirements to store the density matrix as well as Lindblad superoperators. A number of approximation schemes have been developed over time geared towards reducing this difficulty. Some schemes apply perturbation theory \cite{Cirac1992,Reiter2012,Li2014,Haddadfarshi2015} or semi-classical methods \cite{Casteels2017,Lee2012} and are usually limited to specific parameter regimes. 
Interestingly, experimental achievements in increasing coherence times  -- by as much as 6 orders of magnitude for superconducting qubits over the last 20 years \cite{Devoret2013,Wallraff2004} -- further add to the numerical challenges, especially in the context of predicting the long-time asymptotic behavior of quantum systems of interest. For decoherence times vastly exceeding characteristic dynamical time scales associated, e.g., with the drive period, direct integration of the master equation can be both computationally inefficient and vulnerable to numerical rounding errors using standard differential equation solvers. This is especially true for systems with a large Hilbert space \cite{Johansson2012}. 

One approach to obtain the long-time asymptotic solution to the master equation is to employ the Floquet formalism \cite{Kohler1997,Ho1986,Hanggi1998}. However, analytical solutions can usually only be obtained in the adiabatic or high-frequency limit \cite{Bukov2015,Mananga2011,Reimer2018,Dai2016,Restrepo2017}. Furthermore, obtaining the required Floquet basis via matrix diagonalization can be numerically expensive due to the large size of the extended Hilbert space. A method avoiding the switch to the Floquet basis altogether was recently proposed by Hartmann et al.\ \cite{Hartmann2017}. It is based on constructing the Floquet map, i.e., the single-period dissipative propagator of the system, which can be numerically challenging in its own right. By calculating the fixed point of this map, their method resolves the density matrix at stroboscopic instances of time. 

For certain, simple cases of driven open systems, direct numerical integration can be avoided by performing a rotating-frame transformation that eliminates the oscillatory time dependence in the Hamiltonian exactly. In the rotating frame, one can then solve for the nonequilibrium steady-state $\rho_{s}$ which is independent of the initial conditions \cite{Davies1974,Alicki2007} and represents the long-time asymptotic behavior. 
Finding $\rho_s$ amounts to solving a linear system of equations $\mathbb{L}\rho_{s} = 0$, which is generally more efficient than evolving the ODE system to long times, and is not vulnerable to numerical integration errors.  
However, the exact elimination of time dependence is not possible for many systems of interest. One example of interest is the system recently studied by Earnest et al.\ \cite{Earnest2018}: a heavy-fluxonium qubit coupled to a resonator. Direct numerical integration is especially challenging in this case, as the device exhibits a metastable state with lifetimes of up to 8 ms, millions of times longer than the characteristic time scale of the device. 

In this paper we will address this issue by establishing an effective time-independent formalism that approximates the asymptotic solution to the master equation, $\rho_{\infty}(t)$. By adaptively neglecting irrelevant drive terms, we can reduce the system's Hamiltonian to an approximate, effective Hamiltonian that becomes time-independent in an appropriate rotating frame. Such an adaptive rotating-wave approximation (RWA) scheme was previously applied to closed systems in work by Whaley and Light \cite{Whaley1984} and by Einwohner, Wong, and Garrison \cite{Einwohner1976}. 


The structure of our paper is as follows. In Sec.\ \ref{preliminary discussion}, we discuss the general circumstances under which the RWA can lead to a time-independent description in a rotating frame. In Sec.\ \ref{perturbation theory} we then present an iterative scheme which ranks drive terms according to dynamical relevance, adaptively determining the form of the effective Hamiltonian. Sec.\ \ref{application} illustrates applications of the adaptive-RWA scheme, including the simulation of single-tone transmission in the fluxonium-resonator device by Earnest et al.\ \cite{Earnest2018}.  Section \ref{tests} discusses possible limitations of the adaptive-RWA approach. We conclude in Sect.\ \ref{conclusion} and give an outlook on future directions including the extension to multi-periodic Hamiltonians and simulation of two-tone spectroscopy data.

\section{Preliminary Discussion: Frame Transformations} \label{preliminary discussion}  

In a large variety of cases, coherently driven quantum systems are described by a generic time-dependent Hamiltonian of the form
\begin{equation} \label{eq:1}
H(t) = H_{0} + (Ve^{i\omega_{d} t} + \text{h.c.}).
\end{equation} 
Here, $H_{0}$ is the bare system Hamiltonian with eigenstates $\ket{n}$, and $V$ is a system operator that couples to the external drive. As part of the usual rotating-wave approximation (see, e.g., Refs.\ \onlinecite{Whaley1984,Mukamel1976,Larsen1976,Muthu2000}), we assume that the system operator $V$ may be limited to drive terms lowering the system state, i.e., 
\begin{equation}\label{eq:V}
V = \sum_{n < m}V_{nm}\ket{n}\!\bra{m}.
\end{equation}
To account for the fact that the system couples to environmental baths, we describe its open-system dynamics
 by the time-dependent Lindblad master equation \cite{Gorini1976,Lindblad1976,Francesco07}:
\begin{align}\label{eq:2}
   \frac{d\rho (t)}{dt} &= -i \left [H(t),\rho (t) \right] +\sum_{\omega} \gamma_{\omega}\, \mathbb{D}[A_{\omega}]\rho(t).
\end{align}
It describes the interaction of the open system with its environment through a set of collapse operators, $A_{\omega}$, and  associated decoherence rates $\gamma_{\omega}$. Here, $\omega$ denotes differences in system eigenenergies. We have $\omega=0$ for pure dephasing, and $\omega \gtrless 0$ for spontaneous relaxation or thermally-activated excitations, respectively. The dissipation superoperator has the standard form $\mathbb{D}[A_{\omega}]\rho \equiv A_{\omega}  \rho A_{\omega}^\dagger
  - \frac{1}{2} \{   A_{\omega}^\dagger A_{\omega},\rho \}$. Throughout our paper, we will assume that the decoherence channels present are sufficient to guarantee solutions of Eq.\ \eqref{eq:2} to approach a unique, periodic density matrix, independent of the initial state \cite{Yudin2016, Hartmann2017}. This long-time asymptotic behavior or ``Floquet steady-state", $\rho_{\infty}(t)$, is the relevant quantity for the simulation of a number of measurement protocols including transmission and spectroscopy experiments.

In certain situations, a rotating-frame transformation can render the transformed Hamiltonian $h$ (and Lindbladian) time-independent. In this case, the long-time asymptote corresponds to the steady-state solution, $\rho_\infty = \rho_{s}$, obtained from the equation
\begin{equation}\label{eq:3}
   0 = -i  [h,\rho_{s} ] + \sum_{\omega} \gamma_{\omega} \mathbb{D}[A_{\omega}]\rho_{s}.
\end{equation}
Let us inspect under what conditions exact elimination of time dependence can succeed. The rotating-frame transformation is based on a time-dependent unitary matrix, $U(t) = e^{-i\Omega t}$ with generator $\Omega$. For the transformation to eliminate time-dependence in the Hamiltonian, and not introduce time-dependence in the dissipators, we require $[\Omega$, $H_{0}] = 0$. The Hamiltonian thus transforms according to: $H(t) \rightarrow  h(t) = H_{0} - \Omega + U^\dag (t)[ V  e^{i\omega_{d} t} + \text{h.c.}]U(t)$. Since the collapse operators $A_{\omega}$ are eigenoperators of $H_{0}$, the dissipator terms $\mathbb{D}[A_{\omega}]\rho_s$ remain invariant under this transformation. Plugging in Eq.\ \eqref{eq:V} for $V$ and rewriting $\Omega$ in the eigenbasis of $H_{0}$, $\Omega = \sum_{j}\varOmega_{j}\ket{j}\!\bra{j}$ with $\varOmega_{n}$ parametrizing the frame transformation, we observe that the drive terms acquire phase factors: $\ket{n}\!\bra{m} \rightarrow \ket{n}\!\bra{m}e^{i(\varOmega_{n}-\varOmega_{m})t}$. As a result, the rotating-frame Hamiltonian now reads
\begin{equation*}
h(t)=H_{0} - \Omega + \Big(\sum_{n<m}V_{nm}\ket{n}\!\bra{m}e^{i(\varOmega_{n}-\varOmega_{m} + \omega_{d})t} + \text{h.c.}\Big).
\end{equation*}
For $h(t)$ to be time-independent, the constraint $\varOmega_{m} - \varOmega_{n} = \omega_{d}$ must be satisfied for all $n < m$ with $V_{nm}\not=0$. Defining $k_{n} \equiv \varOmega_{n} / \omega_{d}$, we arrive at the central integer constraint
\begin{equation}\label{constraint}
k_{m} - k_{n} = 1.
\end{equation}
Without loss of generality, we can choose all $k_{n}$ to be integers. In conclusion, the possibility to eliminate time dependence exactly hinges upon whether we can assign integers $k_n$ to each system state, such that the integer constraint \eqref{constraint} is satisfied for all drive terms. Let us consider some concrete examples.


If the system is a driven harmonic oscillator, then an $\Omega$ obeying the above integer constraint can be constructed quite easily. The driven-oscillator Hamiltonian $\eqref{eq:1}$ is
\begin{equation}\label{eq:4}
H(t) = \omega_{r}a^\dag a + \zeta (a\,e^{i\omega_{d} t} + \text{h.c.}),
\end{equation}
where $a$ denotes the usual lowering operator for the oscillator with angular frequency $\omega_r$, and $\zeta$ is the drive strength. Following the above notation, this implies $V=\sum_{n=1}^\infty\zeta \sqrt{n} \ket{n-1}\!\bra{n}$. Time dependence is eliminated by setting $\Omega = \omega_{d} a^\dag a = \omega_{d}\sum_{n}n\ket{n}\!\bra{n}$, i.e., $k_{n} = n$ which obviously satisfies the integer constraint for the non-zero drive terms (here, only nearest-neighbor transitions). The transformed Hamiltonian 
\begin{align}\label{eq:5}
h = (\omega_{r} - \omega_{d})a^\dag a + \zeta (a + a^\dag)
\end{align}
is time-independent.

Another example of a system where time dependence can be eliminated exactly is that of a transmon qubit coupled to a resonator: in the limit $E_{J} \gg E_{C}$ only nearest-neighbor qubit transitions appear in the coupling Hamiltonian \cite{koch07}. The system is modeled in terms of an extended Jaynes-Cummings Hamiltonian:
\begin{align}
H(t) = & \, \omega_{r}a^\dag a + \sum_{j}\omega_{j}\ket{j}\!\bra{j} + \sum_{j}g_{j}(a\ket{j+1}\!\bra{j} + \text{h.c.}) \nonumber\\ &+ \zeta (a\, e^{i\omega_{d} t} + \text{h.c.}).
\label{eq:7}
\end{align}
Here, $\ket{j}$ denotes the bare transmon eigenstate with energy $\omega_j$. Due to the nearest-neighbor form of the coupling between resonator and qubit in Eq. $\eqref{eq:7}$, time-dependence can also be eliminated for this system using the generator $\Omega=\omega_{d}(a^\dag a + \sum_{j}j\ket{j}\!\bra{j})$. Expressed in terms of the eigenstates of the generalized Jaynes-Cummings Hamiltonian, this generator reads $\Omega = \omega_{d}\sum_{J,m_J}J\ket{J,m_J}\!\bra{J,m_J}$, where $J$ denotes the combined excitation level of the transmon and resonator, $J = j + n$, and $m_J$ is an integer in the range $0 \leq m_J \leq J$. Each integer (previously denoted $k_n$) is thus given by the total excitation level $J$ for the corresponding dressed state $\ket{J,m_J}$. The transformed time-independent Hamiltonian in the dressed basis is \begin{equation}\label{transmon_transformed}
h = \sum_{J,m_J}(E_{J,m_J}-J\,\omega_d)\ket{J,m_J}\!\bra{J,m_J} + \zeta(a + a^\dag),
\end{equation}
in which $E_{J,m_J}$ are the eigenenergies of the generalized JC Hamiltonian.

For systems with a different structure of non-zero drive terms $V_{nm}$ (e.g., no selection rule limiting the system to nearest-neighbor transitions), satisfying the $k_{m} - k_{n} = 1$ constraint for all $n < m$ may be impossible. This is certainly true for systems consisting of a fluxonium qubit coupled to a resonator, since fluxonium lacks simple selection rules. The oscillatory time dependence in $H(t)$ then cannot be eliminated exactly, no matter the choice of $k_n$. Nevertheless, in the spirit of the RWA, a particular drive term $V_{nm}\ket{n}\!\bra{m}$ may be neglected if it does not significantly affect the system's dynamics. For example, if $V_{nm}$ is very small compared to other drive-term coefficients, or if the drive frequency is far detuned from the energy splitting between $\ket{n}$ and $\ket{m}$, then it may be permissible to neglect  drive term $V_{nm}\ket{n}\!\bra{m}$. 

We thus want to determine whether we are able to approximate the dynamics using an effective Hamiltonian in which a subset of irrelevant drive terms has been neglected, and which becomes time-independent in the appropriate rotating frame. This adaptive RWA would then allow us to extract the long-time asymptotic behavior from a nonequilibrium steady-state.

\section{Adaptive Rotating-Wave Approximation} \label{perturbation theory}

We now develop a systematic scheme to determine whether some of the drive terms can  be neglected, and the problem be reduced to a time-independent one. To assess the importance of each particular drive term, we will consider its contribution to the open-system dynamics as described by the master equation. One common situation leading to negligible influence of a drive term is that of off-resonant driving. For instance, a drive acting on a qubit with drive frequency tuned off resonance relative to the qubit will typically be less effective in inducing Rabi flopping. We will thus seek to distinguish between relevant and irrelevant drive terms, denoting the relevant ones by $V_0$. Once this distinction is established, we may be able to employ an effective Hamiltonian 
\begin{equation}\label{effHam}
H_{\mathrm{eff}}(t) = H_{0} + (V_{0}e^{i\omega_{d} t} + \text{h.c.}),
\end{equation}
in which irrelevant terms are neglected. A key advantage is gained if the remaining drive terms are so simple that a transformation into an appropriate rotating frame eliminates time dependence altogether.


Any method for separating relevant from irrelevant drive terms has to meet two challenges. First, relevance cannot merely be based on energetic resonance conditions, but must also take into account drive strengths, transition matrix elements, as well as the question whether one of the two states involved in a drive term is occupied to begin with. Here, occupation of excited states may arise from other active terms in the drive or be induced thermally. Second, neglecting sub-dominant drive terms only leads to a substantial simplification if it opens up the possibility of a time-independent description by a rotating-frame Hamiltonian 
\begin{equation}
 h = H_{0} - \Omega + (V_{0} + \text{h.c.}).
\end{equation}

\begin{figure}
	\includegraphics[width=0.9\linewidth,trim = 2cm 13.5cm 11.5cm 2cm]{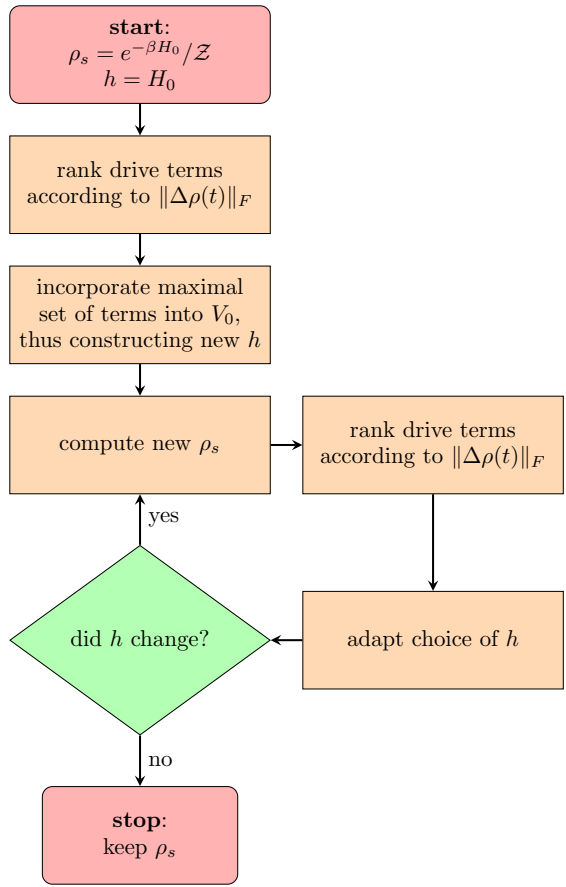}
	\caption{Flowchart for the adaptive-RWA scheme. In this iterative scheme, drive terms are ranked by estimating relevance from magnitude of perturbative corrections to the density matrix. Based on the ranking, a maximal set of drive terms is incorporated into the effective Hamiltonian, allowing for the computation of an approximate time-independent steady-state $\rho_s$ in an appropriate rotating frame.}
	\label{fig:flowchart}
\end{figure}

To address these challenges, we pursue the following strategy (see Fig. \ref{fig:flowchart} for a flowchart summary). We construct $V_{0}$ by  attempting to treat each drive term perturbatively. Specifically, we calculate the perturbative shift of the density matrix induced by individual terms and, thus, establish a relevance ranking among drive terms. Based on this ranking and the goal to enable a time-independent description, a maximal set of terms will be incorporated  into $V_{0}$. Since the relevance of one drive term may depend on the effect of another drive term, we perform multiple iterations of these steps, adaptively changing the terms incorporated into $V_0$ until convergence is reached.   



\subsection{First iteration (bootstrapping)}\label{ssectionA}

To jump-start our iterative scheme, we will initially rank drive terms according to their capacity for steering the system away from the thermal-equilibrium state. In other words, we express the asymptotic solution to the Lindblad master equation $\eqref{eq:2}$ in the form
\begin{equation}\label{eq:9}
\rho_{\infty}(t) = \rho_{s} + \Delta\rho(t),
\end{equation}
where $\rho_{s}=e^{-\beta H_{0}}/\mathcal{Z}$ is the equilibrium state reached in the complete absence of a drive,
\begin{align}
\ 0 = -i \left [H_{0},\rho_{s}  \right] + \sum_{\omega}\gamma_{\omega}\,\mathbb{D}[A_{\omega}]\rho_{s}. \label{eq:10}
\end{align}
The quantities $\beta$ and $\mathcal{Z}$ denote inverse temperature and the partition function, respectively. The correction $\Delta\rho(t)$ reflects the deviation of the system state from equilibrium due to a \textit{single} drive term, $V_{nm}e^{i\omega_{d} t}\ket{n}\!\bra{m} + \text{h.c.}$ Note that $\Delta\rho(t)$ depends on the individual drive term choice. For simplicity, we suppress this dependence on indices $n,\,m$ in our notation. We will take the Frobenius norm of the correction,
\begin{equation}
\|\Delta \rho\|_{F}\equiv \Big(\sum_{i,j}|\Delta\rho_{ij}|^{2}\Big)^{\tfrac{1}{2}},
\end{equation}
which we will use to rank drive term relevance. This is a convenient measure because the time dependence in $\Delta \rho(t)$ will drop out after taking its norm, as we will see below.  

Next, we calculate the corrections $\Delta \rho(t)$ due to each individual drive term in first-order perturbation theory. Upon plugging Eq.\ \eqref{eq:9} and $H=H_0+(V_{nm}e^{i\omega_{d} t}\ket{n}\!\bra{m} + \text{h.c.})$ into the master equation $\eqref{eq:2}$, we can expand in the perturbation $V_{nm}$. The resulting first-order correction obeys the equation
\begin{align}
&\frac{d}{dt}\Delta\rho (t) = -i \left [H_{0},\Delta\rho(t) \right] \label{eq:11}\\\nonumber  
&\, + \sum_{\omega}\gamma_{\omega}\mathbb{D}[A_{\omega}]\Delta\rho(t) - i\left [(V_{nm}\ket{n}\!\bra{m}e^{i\omega_{d} t} + \text{h.c.}),\rho_s  \right].
\end{align}
Note this equation has both a homogeneous solution that depends on initial conditions and a particular solution that depends on the drive term. The asymptotic density matrix, Eq.\ \eqref{eq:9}, does not depend on the initial state, so we seek only the particular solution to this equation. We will solve it by Fourier expanding $\Delta\rho(t) = \sum_{\kappa\in\mathbb{Z}}\varrho_\kappa e^{i\kappa\omega_{d} t}$. Plugging this into Eq. $\eqref{eq:11}$, we obtain equations for the Fourier coefficients $\varrho_\kappa$. Due to the time-dependent phase factors only the coefficients with $\kappa = \pm 1$ are non-zero:
\begin{equation*}
-\omega_{d}\,\varrho_1 = [H_{0},\varrho_1 ] + i\sum_{\omega}\gamma_{\omega}\mathbb{D}[A_{\omega}]\varrho_1 +[V_{nm}\ket{n}\!\bra{m}, \rho_s],
\end{equation*}
and $\varrho_{-1} =  \varrho_1^\dag$.
The only non-zero matrix element of the upper-triangular matrix $\varrho_{1}$ is
\begin{equation}\label{eq:14}
\langle n |\varrho_1 |m\rangle = \frac{V_{nm}(p_m - p_n)}{\omega_{mn}  - \omega_{d} + i(\Gamma_{n} + \Gamma_{m}) / 2}.
\end{equation}
Here, $\omega_{mn}=E_m-E_n$ is the difference between the $m^{\text{th}}$ and $n^{\text{th}}$ eigenenergy of $H_{0}$, $p_n=e^{-\beta E_{n}} / \mathcal{Z}$ is the thermal occupation probability of eigenstate $n$, and $\Gamma_{n}=\sum_{n'} \gamma_{\omega_{nn'}}$ the total decoherence rate of state $n$. The norm $\|\Delta \rho(t)\|_{F}$ is re-expressed in terms of the component $\varrho_{1}$ as
\begin{align}\label{frobnorm}
 \|\Delta \rho(t)\|_{F} = &\sqrt{2}\|\varrho_{1}\|_{F}=\sqrt{2}|\langle n |\varrho_1 |m\rangle|,
 \end{align}
in which, indeed, all time-dependence drops out. For given drive indices $n,\ m$, we thus define the relevance parameter as
\begin{equation}\label{eq:relprm}
 \Delta_{nm} \equiv \|\Delta \rho(t)\|_{F}.
\end{equation}
The relevance parameter $\Delta_{nm}$ characterizes the ability of the drive term to establish coherent oscillations between states $n,\,m$. Inspection of Eq.\ \eqref{eq:14} reveals that multiple factors increase relevance: (i) large transition matrix elements $|V_{nm}|$; (ii) the drive being close to resonance, $\omega_{mn}\approx\omega_d$; (iii) large differences in occupation probabilities between the two involved states $n,\,m$. If both eigenstate populations are thermally suppressed or if they both have similar populations, then the drive term is not as effective at inducing coherent oscillations between the two states and thus the relevance parameter decreases.

All nonzero relevance parameters are now ordered according to magnitude, $\Delta_{n_{1}m_{1}} \geq \Delta_{n_{2}m_{2}} \geq \cdots >0$, into a set $\mathcal{C} = \{\Delta_{n_{1}m_{1}}, \Delta_{n_{2}m_{2}}, \ldots \}$ which provides us with a ranking of the drive terms, see Table \ref{table1}. Based on this, we will next attempt to construct a rotating frame in which the resulting effective Hamiltonian is time-independent and a new steady-state can be obtained.
\begin{table}
\caption{\label{termtable}Drive terms ordered according to the magnitude of the corresponding relevance parameter [Eqs. \eqref{eq:14}-\eqref{eq:relprm}]. 
}
\begin{ruledtabular}
\begin{tabular}{ccc}\label{table1}
\textbf{rank} & \textbf{relevance parameter} & \textbf{drive term}\\
\hline
$1$ (highest) & $\Delta_{n_{1}m_{1}}$ & $V_{n_{1}m_{1}}\ket{n_{1}}\!\bra{m_{1}}+\text{h.c.}$\\
$2$ & $\Delta_{n_{2}m_{2}}$ & $V_{n_{2}m_{2}}\ket{n_{2}}\!\bra{m_{2}}+\text{h.c.}$\\
\vdots & \vdots & \vdots
\end{tabular}
\end{ruledtabular}
\end{table}

\subsection{Determination of the Effective Hamiltonian}\label{ssectionB}
Our goal is to incorporate the maximal set of relevant  drive terms into the effective Hamiltonian, making use of the ranking $\mathcal{C}$ and imposing the integer constraints $k_{m} - k_{n} = 1$ to construct a rotating frame where time dependence is eliminated. To facilitate this, we employ an algorithm similar to the one by Einwohner et al.\ \cite{Einwohner1976}. We represent the drive Hamiltonian as a weighted graph which encodes $V_{0}$ as its maximal zero-cyclic subgraph. While Einwohner et al.\ exclusively consider near-resonant drive terms, we do incorporate lower-ranked off-resonant drive terms whenever possible. The constructed graphs also enable us to read off the selected rotating-frame generator $\Omega$.   

Each nonzero drive term $V_{nm}\ket{n}\!\bra{m}$ (where $m>n$) is graphically depicted by a directed edge connecting the two vertices for states $\ket{n}$ and $\ket{m}$ from left to right. The weight of each edge is set by the corresponding relevance parameter $\Delta_{nm}$. 
Since we wish to track integer constraints \eqref{constraint} throughout the graph, we assign integer labels $k_{n},\, k_m$ to the vertices. Graph edges and vertices are added sequentially, starting with the highest ranked drive term. For a given edge connecting $n,\, m$, there are three possible scenarios for graph construction: (i) neither vertex has been incorporated into the graph yet; (ii) only one has been previously incorporated; (iii) both vertices have already been incorporated. For case (i), we assign the integers $k_{n} = 0$ and $k_{m} = 1$ to the vertices:
\begin{center}
\begin{tikzpicture}[
roundnode/.style={circle, draw=black!60, fill=black!5, very thick, minimum size=10mm},node distance=2cm
]
\node[roundnode,label=below:{$\ket{n}$}]        (0)       {$0$};
\node[roundnode,label=below:{$\ket{m}$}]        (1)       [right=of 0] {$1$};
\draw[ultra thick,->] (0.east) -- (1.west) node[midway, above]{$\Delta_{nm}$};
\end{tikzpicture}
\end{center}
Recall that these integers characterize the generator $\Omega = \omega_{d}\sum_{j}k_{j}\ket{j}\!\bra{j}$ and by choosing $k_{n} = 0$ and $k_{m} = 1$ here ensures the corresponding drive term does not carry a time-dependent phase factor in this rotating frame. For case (ii), we assign an integer to the new vertex, adhering to the integer constraint:
\begin{center}
\begin{tikzpicture}[
roundnode/.style={circle, draw=black!60, fill=black!5, very thick, minimum size=11mm},node distance = 2cm
]
\node[roundnode,label=below:{$\ket{n}$}]        (0)                              {$k_{n}$};
\node[roundnode,label=below:{$\ket{m}$}]        (1)       [right=of 0] {\small $k_{n}$$+$$1$};
\draw[ultra thick,->] (0.east) -- (1.west) node[midway, above]{$\Delta_{nm}$};
\end{tikzpicture}
\end{center}

For case (iii), there are two sub-scenarios. In the first sub-scenario, the two vertices have already been included in the graph, but are in two \textit{disjoint} graph components. Then, the integer of one vertex, along with all other vertices sharing its graph component, must be shifted by some integer $k$ to adhere to the constraint. A concrete example showing how to merge two disconnected graph components is provided in Appendix \ref{appendix A}. While the merging can be accomplished in multiple ways, we show in in Appendix \ref{appendix B} that the resulting graphs only differ by a global integer shift and hence lead to equivalent results.

In the second sub-scenario, both vertices have already been included in the \textit{same} component. In this case, the edge weighted with $\Delta_{nm}$ completes a graph cycle (see Appendix \ref{appendix A} for more details). If this edge connects two vertices with $k_{m} \neq k_{n} + 1$, then we cannot include this drive term in the effective Hamiltonian and we mark the edge by a dashed arrow:
\begin{center}
\begin{tikzpicture}[
roundnode/.style={circle, draw=black!60, fill=black!5, very thick, minimum size=11mm},node distance = 2cm
]
\node[roundnode,label=below:{$\ket{n}$}]        (0)                              {$k_{n}$};
\node[roundnode,label=below:{$\ket{m}$}]        (1)       [right=of 0] {$k_{m}$};
\draw[ultra thick,dashed,->] (0.east) -- (1.west) node[midway, above]{$\Delta_{nm}$};
\end{tikzpicture}
\end{center}
Drive terms marked in this way are neglected in our approximation. (Whether this approximation is good or not depends on whether dashed edges appear for terms with large relevance parameters or are limited to terms with small $\Delta_{nm}$.) 

The above rules are employed iteratively until the full graph has been constructed. The drive terms $V_0$ that will be incorporated in the effective Hamiltonian $ h = H_{0} - \Omega + (V_{0} + \text{h.c.})$ are represented by the subgraph spanned by solid edges (the maximal zero-cyclic subgraph \cite{Einwohner1976}). In this subgraph, the integer constraint $k_{m} = k_{n} + 1$ is satisfied by construction. As a result, the obtained effective rotating-frame Hamiltonian $h$ is time-independent. 

To give a concrete illustration of this scheme, we consider the simplest example where a cycle appears: a driven three-level system with three nonzero drive terms. If the ranking is $\mathcal{C} = \{\Delta_{01}, \Delta_{02}, \Delta_{12}\}$, then the graph is given by
\begin{center}
\begin{tikzpicture}[
roundnode/.style={circle, draw=black!60, fill=black!5, very thick, minimum size=10mm},node distance = 1.9cm
]
\node[roundnode,label=below:{$\ket{0}$}]        (0)                              {0};
\node[roundnode,label=below:{$\ket{1}$}]        (1)       [right=of 0] {1};
\node[roundnode,label=below:{$\ket{2}$}]        (2)       [right=of 1] {1};
\draw[ultra thick,->] (0.east) -- (1.west) node[midway, above]{$\Delta_{01}$};
\draw[ultra thick,dashed,->] (1.east) -- (2.west) node[midway, above]{$\Delta_{12}$};
\draw[ultra thick,->] (0.north east) .. controls +(1,1) and +(-1,1) .. (2.north west) node[midway, above]{$\Delta_{02}$};
\end{tikzpicture}
\end{center} 
The terms given by solid edges, $V_{01}\ket{0}\!\bra{1}$ and $V_{02}\ket{0}\!\bra{2}$, are incorporated into the effective Hamiltonian, while the term $V_{12}\ket{1}\!\bra{2}$ is neglected. By assigning integers for terms in the order determined by the weights in $\mathcal{C}$, we ensure the effective Hamiltonian includes the terms associated with the largest relevance parameters.

\subsection{Subsequent Iterations}\label{ssectionC}

Employing the constructed effective Hamiltonian $h$, we compute the new steady-state $\rho_{s}$ from the master equation
\begin{equation}\label{newsteadystate}
0 = -i \left [h,\rho_{s} \right] + \sum_{\omega}\gamma_{\omega}\,\mathbb{D}[A_{\omega}]\rho_{s}.
\end{equation}
Since bootstrapping bases the relevance of drive terms on the thermal-equilibrium state, the resulting $\rho_s$ may not be a good approximation yet. In subsequent iterations of the adaptive scheme, relevance parameters are re-evaluated based on this new $\rho_s$, thus accounting for the possibility that relevance of drive terms can develop interdependences, especially in cases of multiple (near-)resonant terms.

As before, we consider the effect of each individual drive term $\sim$$V_{nm}$ on the long-time asymptotic behavior of $\rho_\infty(t)=\rho_s + \Delta\rho(t)$. Relevance is based on the magnitude of the deviation from the new steady-state,  $\Delta_{nm}= \|\Delta \rho(t)\|_{F}$.
In the rotating frame, each drive term acquires an additional phase factor, $V_{nm}\ket{n}\!\bra{m}e^{i k_{nm}\omega_{d} t} + \text{h.c.}$, where $k_{nm}\equiv (k_{n} - k_{m} + 1)$, and $k_{n}$ are the previously assigned integers. We solve for $\Delta\rho(t)$ perturbatively, after plugging $\rho_\infty(t)$ and $H = h \pm (V_{nm}\ket{n}\!\bra{m}e^{ik_{nm} \omega_{d} t} + \text{h.c.})$ into the master equation \eqref{eq:2}. Note that the perturbation is added or subtracted, depending on whether it is already part of the current $h$, thus allowing for the possibility that included drive terms may lose relevance in subsequent iterations.

The first-order correction obeys an equation analogous to Eq.\ \eqref{eq:11},
\begin{align}
\frac{d}{dt}\Delta\rho (t) &= -i \left [h,\Delta\rho(t) \right] +
 \sum_{\omega}\gamma_{\omega}\mathbb{D}[A_{\omega}]\Delta\rho(t)
\label{delrho}\\\nonumber
&\quad  \mp i\left [( V_{nm}\ket{n}\!\bra{m}e^{i k_{nm}\omega_{d} t} + \text{h.c.}),\rho_s  \right].
\end{align}
We obtain the particular solution to this equation by Fourier expanding $\Delta\rho(t) = \sum_{\kappa}\varrho_\kappa e^{i\kappa\omega_{d} t}$. Calculating the Fourier components, we find that only the components with $\kappa = \pm k_{nm}$ are non-zero:
\begin{align}\label{fourierTwo}
-k_{nm}\,\omega_{d}\,\varrho_{k_{nm}} &= [h,\varrho_{k_{nm}} ] + i\sum_{\omega}\gamma_{\omega}\,\mathbb{D}[A_{\omega}]\varrho_{k_{nm}} \nonumber\\
&\quad\pm [V_{nm}\ket{n}\!\bra{m}, \rho_s] 
\end{align}
and $\varrho_{-k_{nm}} =  \varrho_{k_{nm}}^\dag$. Solving Eq.\ \eqref{fourierTwo} for $\varrho_{k_{nm}}$ is not as easy as with Eq.\ \eqref{eq:14} in the first iteration, since $h$ and $\rho_{s}$ are now generally non-diagonal matrices. We rewrite Eq.\ \eqref{fourierTwo} more compactly as
\begin{equation} \label{eq:20}
(\mathbb{L}_{0} - ik_{nm}\omega_{d}\openone)\varrho_{k_{nm}} = \mp\mathbb{L}_{nm}\,\rho_{s},
\end{equation}
where the superoperators are defined via $\mathbb{L}_{0}\rho = -i[h,\rho] + \sum_{\omega}\gamma_{\omega}\mathbb{D}[A_{\omega}]\rho$ and $\mathbb{L}_{nm}\rho=-i[V_{nm}\ket{n}\!\bra{m},\rho]$, respectively. Equation \eqref{eq:20} is an inhomogeneous system of linear equations for the $D^2$ components of $\varrho_{k_{nm}}$. 

In solving Eq.\ \eqref{eq:20}, we distinguish two different cases: if $k_{nm} \neq 0$, then the superoperator $\mathbb{L}_{0} - i\,k_{nm}\omega_{d}\openone$ is invertible; if $k_{nm} = 0$, then it is not invertible. To see this, note that, by assumption,  $\mathbb{L}_{0}$ has no purely imaginary eigenvalues. [Recall that we are requiring decoherence channels sufficient to guarantee a unique steady-state given by Eq.\ \eqref{newsteadystate}, $\mathbb{L}_{0}\rho_{s} = 0$.] Since $\det(\mathbb{L}_{0} - i\,k_{nm}\omega_{d}\openone)=0$ if and only if $i\,k_{nm}\omega_{d}$ is an eigenvalue of $\mathbb{L}_{0}$, we can invert $\mathbb{L}_{0} - i\,k_{nm}\omega_{d}\openone$ for $k_{nm} \neq 0$.  For $k_{nm} = 0$  the superoperator $\mathbb{L}_{0} - i\,k_{nm}\omega_{d}\openone=\mathbb{L}_{0}$ is singular. In this case there is an infinite number of solutions, obtained by shifting $\varrho_{k_{nm}}$ by some multiple $c$  of the steady-state, $\varrho_{k_{nm}} \rightarrow \varrho_{k_{nm}} + c\,\rho_{s}$. We can compute $\varrho_{k_{nm}}$ utilizing the Moore-Penrose pseudoinverse $\mathbb{L}_{0}^{+}$ \cite{Ben2003} and shifting the result to render it traceless. Since the pseudoinverse reduces to the standard inverse when the matrix is invertible, we can express the solution in general as
\begin{equation} \label{fouriercomponenttwo}
\varrho_{k_{nm}} = \mp(\mathbb{L}_{0} - i\,k_{nm}\omega_{d}\openone)^{+}\,\mathbb{L}_{nm}\,\rho_{s}.
\end{equation}
Instead of computing the pseudoinverse, one may alternatively employ an efficient least-squares method in which the norm $\|(\mathbb{L}_{0} - i\,k_{nm}\omega_{d}\openone)\varrho_{k_{nm}} \pm\mathbb{L}_{nm}\,\rho_{s} \|_{F}$ is minimized.


As before, we find that application of the Frobenius norm renders the relevance parameter time-independent:
\begin{equation}\label{relparamtwo}
\Delta_{nm} = \|\Delta \rho(t)\|_{F} = \sqrt{2}\|\varrho_{k_{nm}}\|_{F}. 
\end{equation}
The updated relevance parameters are next employed in the graphical scheme of Section \ref{ssectionB} to identify the maximal zero-cyclic subgraph, yielding another new effective Hamiltonian $h$. This iterative scheme is repeated for as long as re-evaluating relevance parameters causes $h$ to change (or until a maximum iteration number is exceeded, indicating rare cases when the method breaks down),
\begin{align}\label{adapt}
\rho^{(0)}_s = e^{-\beta H_{0}}/\mathcal{Z} \;&\rightarrow\; \rho^{(1)}_{s} \rightarrow \rho^{(2)}_{s} \rightarrow \dots \rightarrow \rho_{s} \\\nonumber
h^{(0)} = H_{0} \;&\rightarrow\; h^{(1)} \rightarrow h^{(2)} \rightarrow \dots \rightarrow h.
\end{align}
Here, superscripts enumerate the iterative steps (suppressed in our notation above).

In summary, this iterative scheme adaptively incorporates the most relevant drive terms, and takes into account the maximal set of sub-dominant drive terms. In the next sections we will illustrate the power of the method by applying it to single-tone transmission spectroscopy in a system with metastability, and discuss possible limitations based on a simple three-level system  example.


\begin{figure*}
  \includegraphics[width=0.9\linewidth]{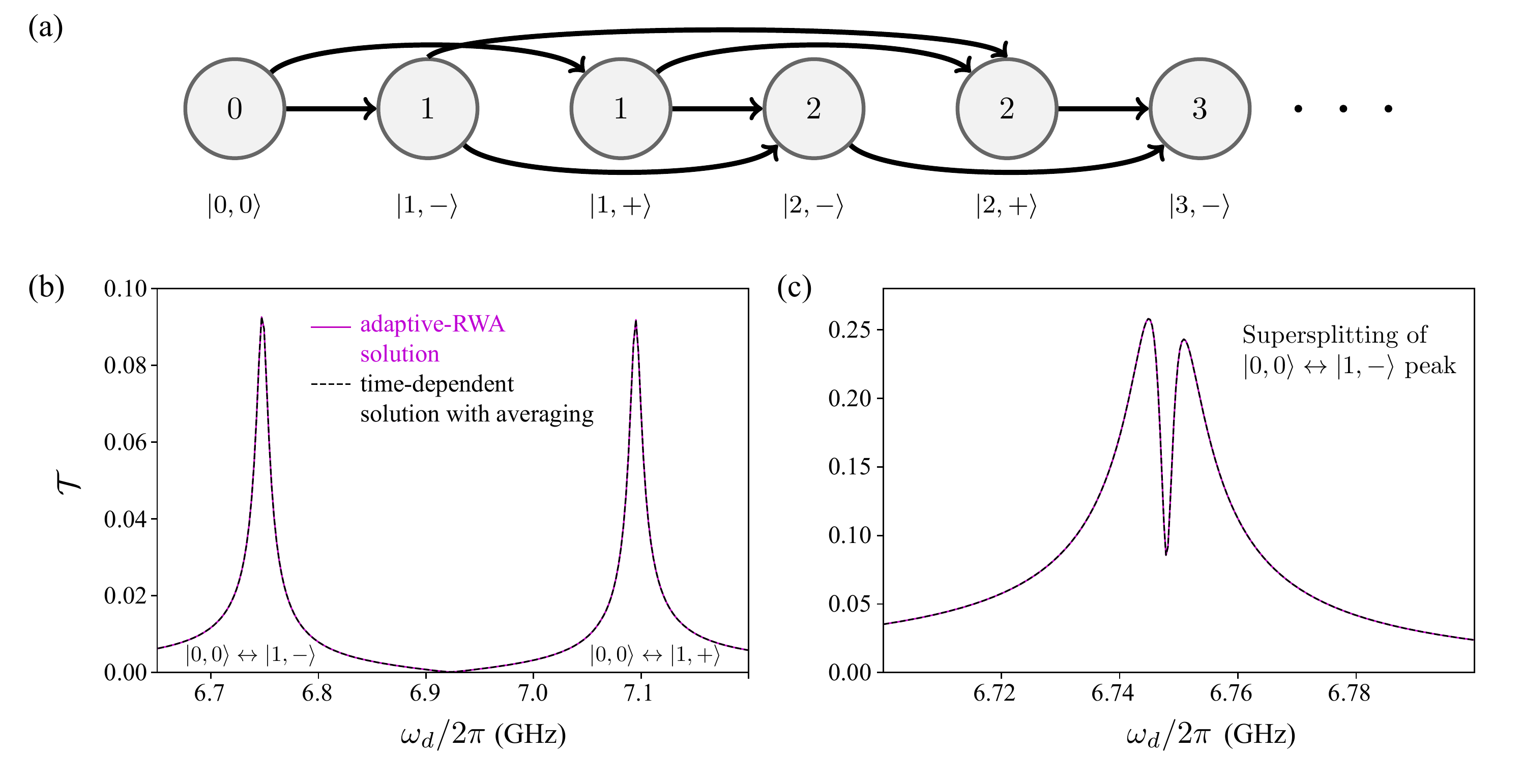}\\[-4mm]
  \caption{Adaptive-RWA calculation of transmission $\mathcal{T}$ vs.\ drive frequency in driven transmon-resonator system. \textbf{(a)} Adaptively obtained graph for weak drive at $\omega_{d} = \omega_{r}$. 
Absence of dashed edges confirms that all drive terms are incorporated, hence the result is exact.
\textbf{(b)} Comparison of exact transmission with results from adaptive RWA, showing excellent agreement. Since transmon and resonator are set to resonance, $\mathcal{T}$ exhibits the usual vacuum Rabi peaks, arising from transitions between $\ket{0,0}$ and $\ket{1,\pm}$.
\textbf{(c)} For a ten-fold increase in drive power, each Rabi peak supersplits \cite{Bishop08}. Exact solution and adaptive-RWA results continue to match perfectly.  
\label{fig:transmon}}
\end{figure*}

\section{Application: Single-Tone Spectroscopy} \label{application}

We illustrate application of the adaptive-RWA method to the calculation of single-tone transmission data for two different circuit-QED systems. First, we show that the scheme reproduces the exact steady-state solution for the simple system of a transmon qubit coupled to a resonator. Second, we simulate recent transmission measurements of a heavy-fluxonium circuit-QED device \cite{Earnest2018}, in which the presence of long-lived metastable states makes ordinary time-dependent simulations particularly challenging. 

In conventional single-tone experiments, transmission of a coherent drive tone through the resonator is probed and utilized to determine the dispersively shifted resonator frequency, or detect the vacuum Rabi splitting, depending on whether the qubit is tuned out of or into resonance. Transmission data for the oscillatory voltage signal is typically averaged over many periods, after transients have died out. In terms of the field quadratures $I = V_{p}\langle a + a^\dag \rangle$ and $Q = V_{p}\,i\langle a^\dag - a \rangle$, where $V_{p}$ is the peak voltage, we express the transmission amplitude as
\begin{equation}\label{eq:22}
A(t) = V_{p}\sqrt{I^{2}+Q^{2}} = 2V_{p}\,|\langle a \rangle| = 2V_{p}\,|\Tr[a\rho(t)]|.
\end{equation}
The averaged transmitted power is thus proportional to $\overline{| \langle a \rangle |^{2}} \equiv \mathcal{T}^2$ where time-averaging is performed on the long-time asymptote $\rho_\infty(t)$. The adaptive-RWA scheme allows us to calculate resonator transmission based on an effective rotating-frame steady state, $\mathcal{T} = |\Tr(a\rho_{s})|$, instead of calculating $\rho_\infty(t)$ numerically by integrating the master equation up to sufficiently long times.

\subsection{Transmon Qubit Coupled to Resonator}\label{transmon}
We first confirm that the adaptive-RWA calculation returns exact results whenever time-dependence can be fully eliminated in an appropriate rotating frame. This situation is realized for the simple example of a system consisting of a transmon and a resonator, as discussed in Sec.\ \ref{preliminary discussion}. Recall that the transmon states form a weakly anharmonic ladder in which the resonator coupling only allows for transitions among nearest-neighbor transmon levels. We expect the adaptive scheme to find the appropriate rotating frame and yield transmission data identical to those from the exact solution.

Previously, we expressed the dressed transmon-resonator eigenstates as $\ket{J,m_J}$ where $J$ is the total excitation level and $m_J$ is an integer in the range $0\leq m_J\leq J$. If the drive strength is not too strong, we can approximate the transmon as a two-level system, and if the qubit and resonator are on-resonance, we can find expressions for the dressed states in terms of the bare states $\ket{n,j}$ with $n$ and $j$ as the resonator and transmon levels respectively. These expressions are $\ket{J,\pm} = (\ket{n=J,j=0} \pm \ket{n=J-1,j=1})/\sqrt{2}$ \cite{Bishop08}, hence the generator is given as $\Omega = \omega_{d}\sum_{J,\pm}J\ket{J,\pm}\!\bra{J,\pm}$ (note that there are only two possible values for $m_J$ here). Applying our adaptive scheme for arbitrary drive frequency, we expect a graph consistent with this $\Omega$, and an effective Hamiltonian that is composed of every non-zero drive term. In Fig.\ \ref{fig:transmon}(a) we show the graph the adaptive scheme converges to.  For the example of driving at the resonator frequency, $\omega_d=\omega_{r}$, the ranking for the first few terms is
\begin{align}
\mathcal{C} = \{\Delta_{01}, \Delta_{02},\Delta_{24}, \Delta_{13}, \Delta_{46}, \Delta_{35},\ldots\},
\end{align}
where the subscripts indicate the energy level of each dressed state [see Fig.\ \ref{fig:transmon}(a)]. We emphasize that the adaptive scheme does not neglect any drive terms in this special case.
 
Figure \ref{fig:transmon} compares between transmission results obtained from the adaptive-RWA scheme and those calculated from the time-dependent master equation after averaging. Since the transmon is placed on resonance, the transmission curve exhibits the characteristic vacuum-Rabi peaks [Fig.\ \ref{fig:transmon}(a)]. For increased drive strength, each vacuum-Rabi peak supersplits [Fig.\ \ref{fig:transmon}(c)] \cite{Bishop08}. Exact results (here obtained from averaging the lab-frame time-dependent solution) and results from the adaptive scheme are in perfect agreement. This confirms that the scheme correctly selects the relevant drive terms and detects the rotating frame in which the effective Hamiltonian is time independent.

\begin{figure*}
\includegraphics[width=\linewidth]{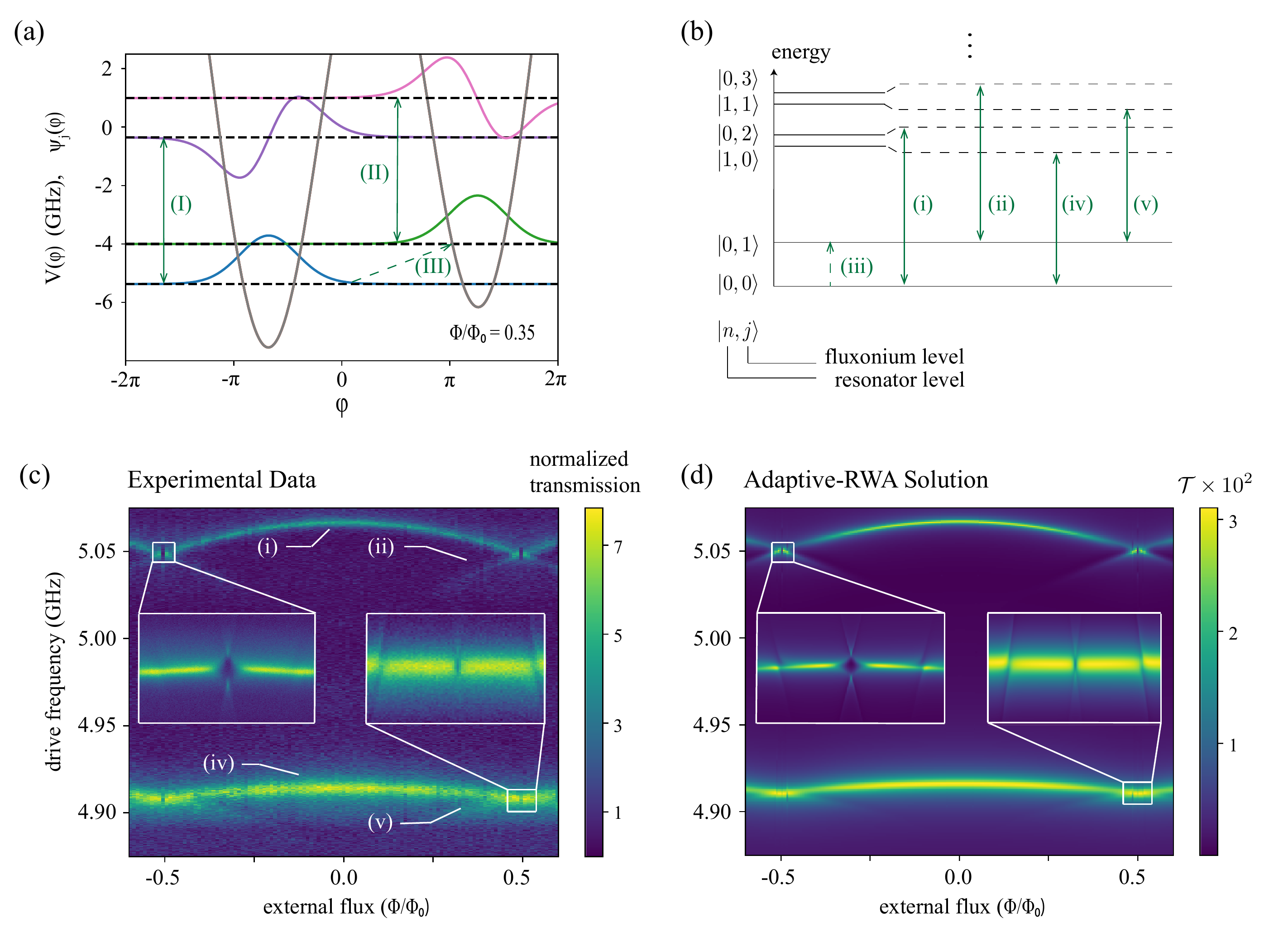} \caption{Heavy-fluxonium spectrum and comparison between experimental transmission data and adaptive-RWA calculation. \textbf{(a)} Relevant heavy-fluxonium states and corresponding transitions ($\Phi = 0.35\Phi_{0}$). Transition I is a plasmon (intra-well) transition between fluxonium states $\ket{0}\leftrightarrow\ket{2}$. Plasmon transition II between $\ket{1}\leftrightarrow\ket{3}$ only occurs if one of these states is thermally excited, e.g., through  fluxon transition III. \textbf{(b)} Overview of dressed energy levels and transitions. 
Up to photon dressing, transitions (i)--(iii) correspond to the ones in (a). \textbf{(c)} Experimental transmission amplitude normalized to high powers (in decibels) in the single-tone spectroscopy experiment \cite{Earnest2018}. The labeled resonances correspond to the transitions defined in (b). \textbf{(d)} Transmission amplitudes $\mathcal{T}=|\Tr(a\rho_{s})|$ obtained from the adaptive-RWA scheme showing excellent agreement with the experimental data.
\label{fig:fluxonium}}
\end{figure*}

\subsection{Heavy-Fluxonium Qubit Coupled to Resonator}\label{fluxonium}
The adaptive-RWA scheme is most useful in situations where time dependence \emph{cannot} be eliminated exactly. We will demonstrate this for a quite recent and promising addition to the family of circuit-QED devices: a heavy-fluxonium qubit coupled to a resonator. 
Again, we focus on the transmission amplitude when a drive is acting on the resonator, and employ the adaptive-RWA algorithm. The Hamiltonian of this system is given by 
\begin{align}\label{eq:23}
&H(t) = \omega_{r}\,a^\dag a + \sum_{j}E_{j}\ket{j}\!\bra{j} \\\nonumber &\quad+ g\sum_{j,j'}\big(\bok{j}{N}{j'}\ket{j}\!\bra{j'}a + \text{h.c.}\big) + \zeta (a\,e^{i\omega_d t} + \text{h.c.})
\end{align}
see, e.g., Ref.\ \cite{Zhu13}. Here, bare fluxonium states have energies $E_j$ and are denoted by $\ket{j}$, $N$ is the fluxonium charge operator, and $\zeta$ the drive strength. This generalized Jaynes-Cummings Hamiltonian differs from the analogous Eq.\ \eqref{eq:7}: as opposed to the transmon case,  fluxonium charge matrix elements are not subject to nearest-neighbor selection rules, so $\bok{j}{N}{j'}$ are generally nonzero for all $j,\,j'$. Accordingly, time-dependence cannot be removed exactly by any rotating-frame transformation.

The experiment by Earnest et al.\ \cite{Earnest2018} uses the heavy fluxonium to realize a $\Lambda$ system with a metastable state featuring lifetimes of up to $8$ ms. Figure \ref{fig:fluxonium}(a) depicts the fluxonium wave functions and potential-well structure for a select magnetic flux of $\Phi = 0.35\Phi_{0}$. Device parameters in the experiment were tuned such that the intra-well (plasmon) energy splitting, $E_{2} - E_{0} \equiv \omega_{20}$, was nearly degenerate with the resonator frequency $\omega_{r}$. This results in strong hybridization of resonator and plasmon modes, rendering the single-tone transmission data richer than usual. In addition, the long dwell-times in the metastable state render rare thermal-excitation processes relevant for the device's long-time dynamics. Indeed, fingerprints of this interplay between metastability and thermal excitations are observed in the form of anomalous peaks in the transmission data which we will discuss in detail next.

Figures \ref{fig:fluxonium}(c) and (d) show experimental data and adaptive-RWA calculations of the transmission (color-coded) as a function of external magnetic flux $\Phi$ and frequency $\omega_d/2\pi$ of the applied drive. The selected frequency range spans the region near $\omega_{r}$ and $\omega_{20}$ to capture the transmission peaks arising from photon excitations of the resonator, transition (iv), and dressed plasmon oscillations, transition (i) [see Fig.\ \ref{fig:fluxonium}(b) for labeling of dressed-state transitions]. The latter plasmon resonance is ordinarily not visible in single-tone transmission experiments when the qubit is coupled dispersively, but can be observed here because the left plasmon transition I, $\ket{0} \leftrightarrow \ket{2}$, is only weakly detuned from the resonator, $\omega_{20}-\omega_r\sim g$. Insets in Figs. \ref{fig:fluxonium}(c) and (d) display the more intricate structure of resonances and avoided crossings in the region near half-integer flux, and confirm the very good agreement between experimental data and our adaptive-RWA results. 

\begin{figure}
	\includegraphics[width=0.9\linewidth, trim = 4.6cm 20cm 4.1cm 5cm]{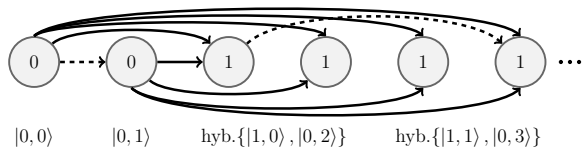}
	\centering
	\caption{Graph for the driven fluxonium-resonator system ($\Phi = 0.35\Phi_{0}$,  $\omega_{d} = \omega_{31}$). Adaptive selection of relevant drive terms depends on $\Phi$ and $\omega_{d}$, and is critical for accurate computation of transmission peaks. Dashed edges mark neglected drive terms with very small relevance parameters due to exponential suppression of charge matrix elements and/or drive detuning. 	\label{fig:fluxonium_diagram}}
\end{figure}

While thermal excitation events populating the metastable state $\ket{1}$ remain rare at a temperature of $30$ mK consistent with experimental conditions, the occupation probability for $\ket{1}$ can nonetheless become significant due to its exceedingly long lifetime. This gives rise to an anomalous transmission peak associated with the dressed transition (ii), visible in both experimental data and simulation. As seen in the graph of Fig.~\ref{fig:fluxonium_diagram}, for a drive frequency $\omega_d=\omega_{31}$ the adaptive-RWA algorithm properly includes the drive terms that induce the dressed-plasmon transitions (ii) in the right potential well, capturing their relevance due to thermal excitations. The time scales for multiple competing thermal-excitation channels vary between milliseconds and seconds, the latter applying to the direct  $\ket{0} \rightarrow \ket{1}$ transition. The resulting vast span of time scales, ranging the from the nanosecond drive period to millisecond excitation times, makes brute-force time evolution and averaging a disadvantageous strategy for numerical simulation.

It is worth noting that numerical integration of the lab-frame master equation does not merely face computational efficiency issues with the excessive integration time in the case of long-lived qubit states, but can also run into serious difficulties due to accumulation of numerical errors. Using standard integrators, we encountered such issues that prevented us from obtaining reliable transmission values from a brute-force time evolution. The adaptive-RWA scheme eliminates this challenge and successfully reproduces the thermally-activated transmission resonances.
  
\begin{figure}
	\includegraphics[width=\linewidth]{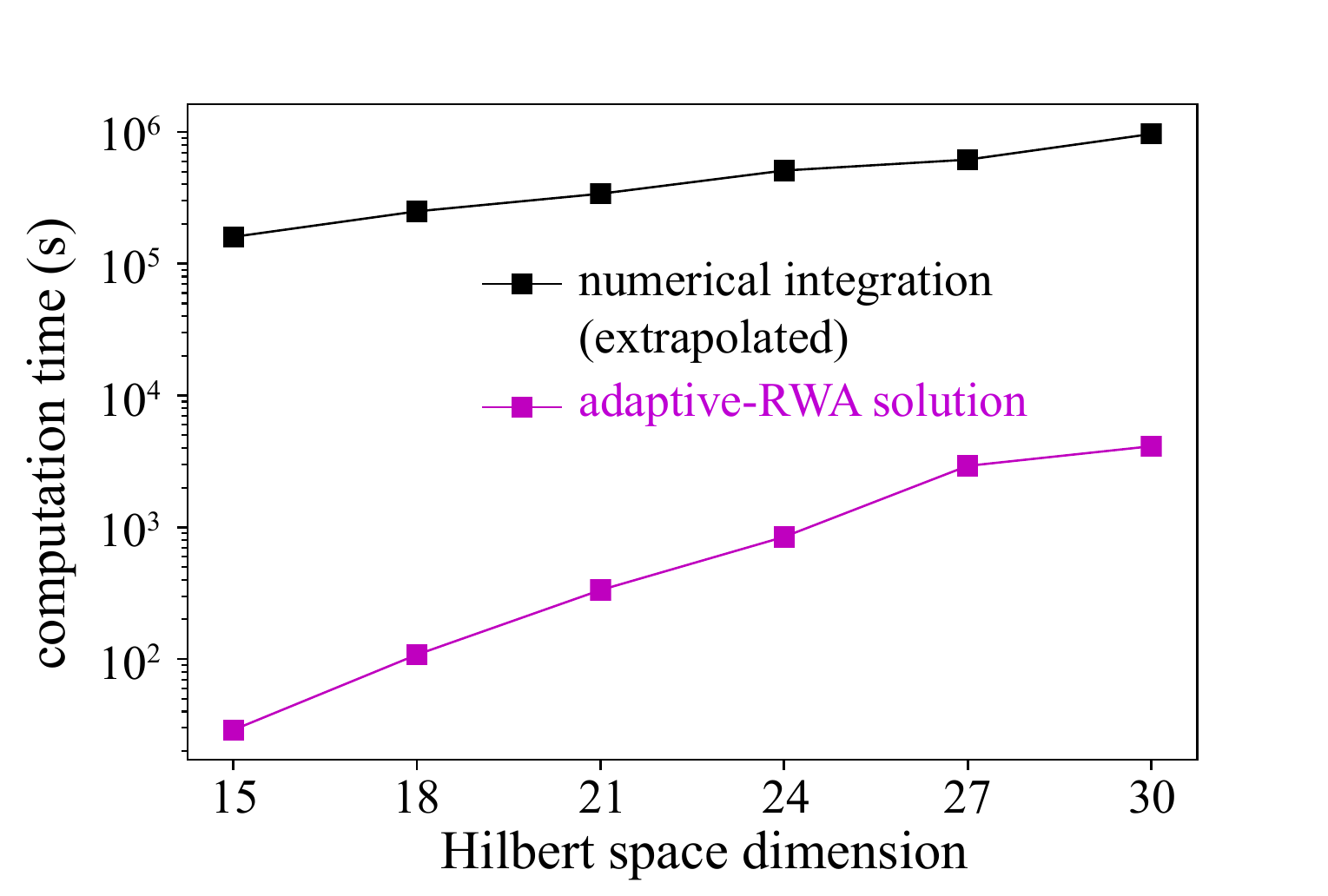}
	\caption{Comparison of computation times using either direct numerical integration of the master equation or the adaptive-RWA scheme, for the example of the heavy fluxonium-resonator system. Computation time using the adaptive-RWA algorithm is orders of magnitudes shorter for moderate Hilbert space dimensions. (Truncation at dimension 15 was sufficient for our chosen parameters of $T=30$ mK, $\Phi=0.35\Phi_{0}$, $\omega_d/2\pi=5.039$ GHz, and $\zeta/2\pi=100$ kHz.) 
    \label{fig:comp_time}}
\end{figure}

A comparison of the required computation time clearly shows the advantage of the adaptive-RWA scheme over the direct numerical integration of the master equation [Fig.\ \ref{fig:comp_time}]. We estimate the computation time for direct numerical integration by extrapolation: the master equation was first integrated numerically over a time interval of $0.5\,\mu\text{s}$, and the required computation time then scaled up for the intended time interval of $5\,$ms -- an appropriate time given the relevance of rare thermal excitations and lifetimes of the metastable state. Figure \ref{fig:comp_time} illustrates how the adaptive-RWA scheme cuts down computation time by a factor of $10^3$ or more in this example. (See App.\ \ref{appendixC} for a computational-cost comparison with the Floquet-map method \cite{Hartmann2017}.)

\section{Limitations of the adaptive-RWA scheme \label{tests}}

The adaptive-RWA scheme is applicable to a broad range of driven open quantum systems. The scheme may fail, however, in special situations where multiple drive terms are similarly relevant and prevent construction of a zero-cyclic graph.  In the following, we discuss this limitation of the adaptive-RWA scheme in the simplest possible context: a driven three-level system in which all drive terms have comparable relevance parameters. 

The Hamiltonian $H(t) = H_{0} + (Ve^{i\omega_{d} t} + \text{h.c.})$ of the driven three-level system consists of $H_{0} = \sum_{n=0}^{2}E_{n}\ket{n}\!\bra{n}$ for the three eigenstates and 
\begin{equation}\label{three_level_V}
V = V_{01}\ket{0}\!\bra{1} +  V_{02}\ket{0}\!\bra{2} + V_{12}\ket{1}\!\bra{2}
\end{equation}
describing the drive terms. The adaptive RWA will succeed, here, if one of these drive terms has low relevance compared to the other two and can be neglected.

\begin{figure}
	\includegraphics[width=0.9\linewidth,trim = 6.5cm 21cm 6.5cm 4.5cm]{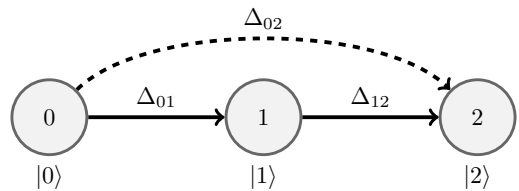}
	\caption{Graph for the three-level system with identical level splittings when driven on resonance, $\omega_{d} \approx \omega_{0}$. Since the $\ket{0}\to\ket{2}$ transition is off-resonant, $\Delta_{02}$ will typically be much smaller than the other relevance parameters, and the adaptive-RWA scheme succeeds.} 
	\label{fig:threelevel_diagram}
\end{figure}

\begin{figure*}
\includegraphics[width=0.9\linewidth]{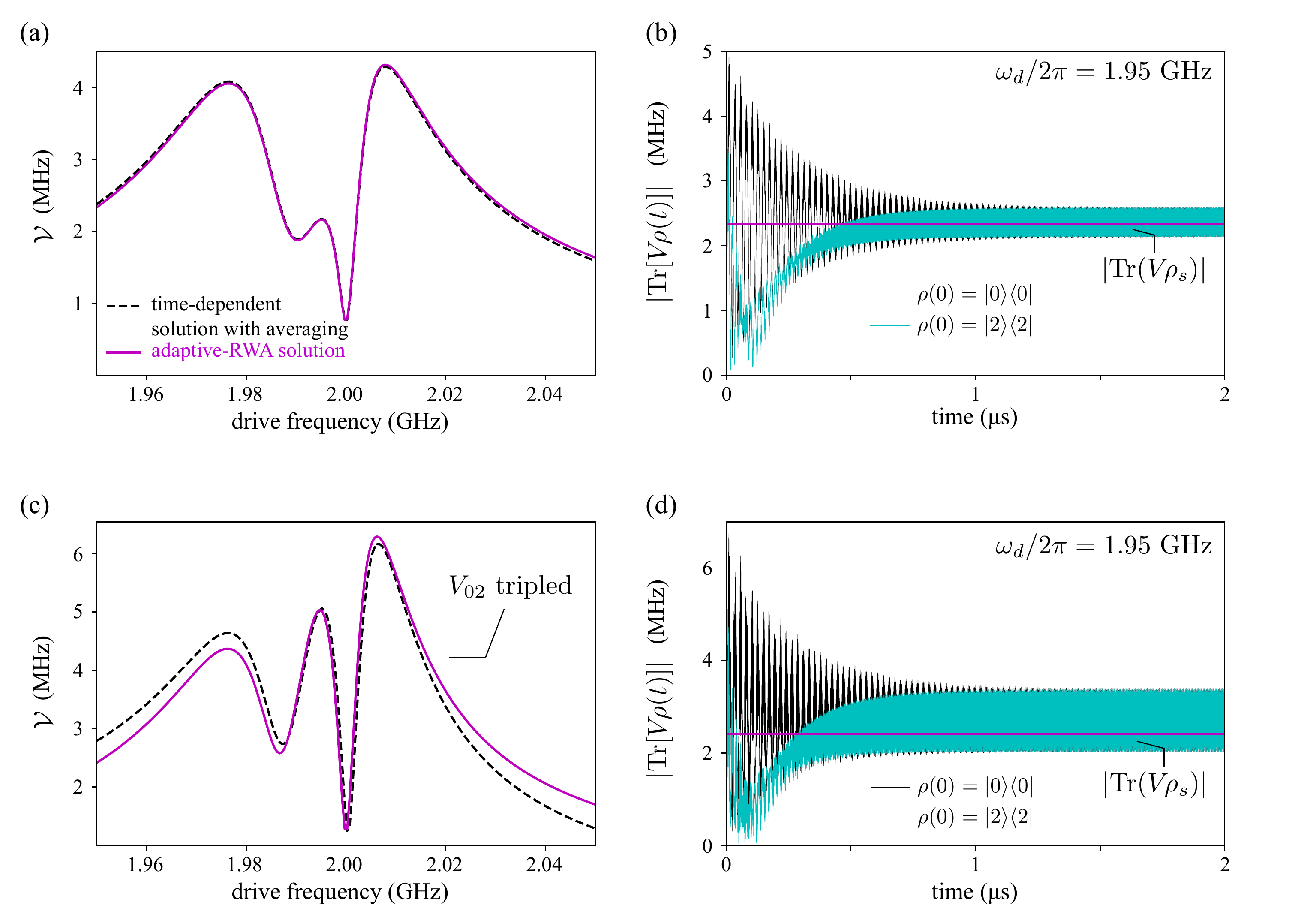}
\caption{Success and breakdown of the adaptive-RWA scheme in the driven three-level system. \textbf{(a)} Comparison of $\mathcal{V}=|\Tr (V\!\rho_s)|$ with the numerically exact result obtained by time-averaging the long-time dynamics. For drive terms $V_{mn}$ of equal magnitude, the off-resonant drive term $V_{02}\ket{0}\!\bra{2}$ has low relevance and can be neglected, thus leading to good agreement between approximation and exact solution. \textbf{(b)} The time-dependent signal for (a) at the frequency $1.95$ GHz, plotted for two different initial conditions. The long-time asymptotes are seen to match the adaptive-RWA result. \textbf{(c)} When $V_{02}$ is tripled in magnitude, this drive term becomes relevant despite being off-resonant and deviations between adaptive-RWA and exact solution become visible. \textbf{(d)} The equivalent time-dependent signal as in (b), with $V_{02}$  tripled.\label{fig:threelevelsystem} }
\end{figure*}

For example, suppose that the the energy-level splittings are nearly identical, $E_{1}-E_{0} \approx E_{2}-E_{1} \equiv \omega_{0}$, and that the drive matrix elements $V_{mn}$ all have the same order of magnitude.
If the system is resonantly driven with frequency $\omega_{d} \approx \omega_{0}$, then the dynamics will be dominated by transitions induced by the drive terms $V_{01}\ket{0}\!\bra{1}$ and $V_{12}\ket{1}\!\bra{2}$. The adaptive-RWA scheme will yield the graph shown in Fig.\ \ref{fig:threelevel_diagram}, based on the ranking $\Delta_{01}\approx\Delta_{12}\gg\Delta_{02}$. Here, $V_{02}\ket{0}\!\bra{2}$ has significantly lower relevance since the $\ket{0}\to\ket{2}$ transition is off-resonant. The resulting effective Hamiltonian in the appropriate rotating frame is given by
\begin{align}\label{three_level_h}
h &=  \,H_0 - \omega_d(\ket{1}\!\bra{1} + \ket{2}\!\bra{2}) \\\nonumber &\quad\; + (V_{01}\ket{0}\!\bra{1} + V_{12}\ket{1}\!\bra{2} + \text{h.c.}).
\end{align}
Here, adaptive-RWA results are good approximations to the asymptotic long-time behavior $\rho_\infty(t)$ of the system.

As an example observable, we calculate $\mathcal{V}\equiv|\Tr ( V \rho_s)|$ for drive frequencies near $\omega_{0}$ -- a quantity similar to the transmission signal $\mathcal{T}$ calculated in the previous section. As expected, Fig.\ \ref{fig:threelevelsystem}(a) shows a supersplit resonance peak, and adaptive-RWA results are in good agreement with the exact solution based on time-averaging $|\Tr [V \rho_\infty(t)]|$. This time-dependent signal is shown explicitly in Fig.\ \ref{fig:threelevelsystem}(b) for two different initial states -- illustrating how the system first passes through a transient phase and then reaches its asymptotic behavior, whose time-average is in agreement with the adaptive-RWA solution.


Breakdown of the adaptive RWA occurs if we raise the relevance of the $V_{02}$ drive term: as $\Delta_{02}$ approaches the magnitude of the other relevance parameters, the corresponding drive term cannot be safely neglected.
 Indeed, if we triple the magnitude of $V_{02}$,  then deviations between the adaptive-RWA solution and the exact result become clearly visible [see Fig.\ \ref{fig:threelevelsystem}(c)].These deviations are likewise reflected in Fig.\ \ref{fig:threelevelsystem}(d), showing that the adaptive-RWA solution does not accurately match the actual long-time asymptotics. As expected, deviations from the exact solution diminish for drive frequencies around $\omega_{0}/2\pi=2$ GHz, i.e., when the system is driven on resonance. In a particular pathological case, $\Delta_{02}$ could become so large that the iterative scheme would not converge. In our experience, such cases are rare and do not naturally occur in common driven circuit-QED systems.

\section{Conclusions and Outlook\label{conclusion}} 

In this paper, we have presented the adaptive-RWA scheme: a numerical method for driven open quantum systems that approximates the asymptotic long-time solution to the master equation by a nonequilibrium steady-state in an adaptively selected rotating frame. By iteratively determining which drive terms in the Hamiltonian are most relevant to the dynamics, the algorithm chooses an effective Hamiltonian including a maximal set of relevant drive terms. Each iteration involves solving an inhomogeneous set of linear equations, and avoids the need to numerically solve the system of ODEs tracking the system dynamics. Adaptive-RWA computations can dramatically improve efficiency over direct numerical integration, particularly when decoherence time scales are as long as those achieved in recent circuit-QED experiments.

We have illustrated applications of the adaptive-RWA scheme to coupled transmon-resonator and fluxonium-resonator systems. We have seen that the adaptive-RWA results reproduce transmission observed in single-tone spectroscopy experiments for heavy-fluxonium done by Earnest et al.\ \cite{Earnest2018}, including the appearance of anomalous, thermally-activated transmission resonances. In general, the adaptive-RWA method is useful for a wide class of driven open quantum systems that do not allow for exact elimination of time dependence within some appropriate rotating frame. The adaptive-RWA scheme proves particularly beneficial in systems with large $T_{1}$ and $T_{2}$ times which make explicit numerical calculation of the long-time asymptotic behavior challenging.


In the future, we plan to extend this adaptive scheme to multi-tone drives, enabling the simulation of two-tone spectroscopy experiments. In the multi-tone case, additional care must be taken when considering the rotating-frame transformation and the effective Hamiltonian due to additional constraints for eliminating time dependence and graph-construction rules. Investigation of multi-tone driving with the adaptive-RWA method offers exciting prospects for studying future experimental systems. Finally, we note that the calculation of relevance parameters from first-order perturbation theory does not account for the occurrence of two-photon transitions. Extending the scheme to higher orders will therefore prove fruitful in situations with larger drive strengths.

\begin{acknowledgments}
We thank Peter Groszkowski for valuable discussions.  This research was supported by the Army Research Office through Grant No.\ W911NF-15-1-0421 and by the NSF Graduate Research Fellowship Program through grant No.\ DGE-1144082. 
\end{acknowledgments}

\appendix

\section{Graph Construction Algorithm \label{appendix A}} 

This appendix details graph construction for an example system that requires merging of two graph components, as previously mentioned in Sec.\ \ref{ssectionB}.  

Suppose the ranking for this example is given by 
\begin{align}
\mathcal{C} = \{\Delta_{01}, \Delta_{23}, \Delta_{45}, \Delta_{04}, \Delta_{13}, \Delta_{24}, \Delta_{34}\}.
\end{align}
Following this ranking, graph construction starts by establishing edges for $V_{01}\ket{0}\!\bra{1}$, $V_{23}\ket{2}\!\bra{3}$, and $V_{45}\ket{4}\!\bra{5}$, leading to three disconnected graph components:
\begin{center} 
\begin{tikzpicture}[
roundnode/.style={circle, draw=black!60, fill=black!5, very thick, minimum size=8mm},
node distance=7.5mm,]
\node[roundnode,label=below:{$\ket{0}$}]        (0)                              {0};
\node[roundnode,label=below:{$\ket{1}$}]        (1)       [right=of 0] {1};
\node[roundnode,label=below:{$\ket{2}$}]		(2)       [right=of 1] {0};
\node[roundnode,label=below:{$\ket{3}$}]        (3)       [right=of 2] {1};
\node[roundnode,label=below:{$\ket{4}$}]        (4)       [right=of 3] {0};
\node[roundnode,label=below:{$\ket{5}$}]		(5)       [right=of 4] {1};
\draw[ultra thick,->] (0.east) -- (1.west)  node[midway, above]{$\Delta_{01}$};
\draw[ultra thick,->] (2.east) -- (3.west)  node[midway, above]{$\Delta_{23}$};
\draw[ultra thick,->] (4.east) -- (5.west)  node[midway, above]{$\Delta_{45}$};
\end{tikzpicture}
\end{center}
The relevance ranking prompts for inclusion of $V_{04}\ket{0}\!\bra{4}$, next, connecting states $\ket{0}$ and $\ket{4}$. This requires merging of two separate graph components, done by shifting \textit{all integers in one component} such that the two states in question, here $\ket{0}$ and $\ket{4}$, can be linked by a solid edge satisfying the integer constraint $k_4-k_0=1$. We have the choice of either down-shifting the component containing $\ket{0}$, or up-shifting the component containing $\ket{4}$. As shown in App.\ \ref{appendix B}, the resulting graphs are always equivalent. 
Choosing to up-shift the right-most graph component by $+1$, we obtain
\begin{center} 
\begin{tikzpicture}[
roundnode/.style={circle, draw=black!60, fill=black!5, very thick, minimum size=8mm},
node distance=7mm,]
\node[roundnode,label=below:{$\ket{0}$}]        (0)                              {0};
\node[roundnode,label=below:{$\ket{1}$}]        (1)       [right=of 0] {1};
\node[roundnode,label=below:{$\ket{2}$}]		(2)       [right=of 1] {0};
\node[roundnode,label=below:{$\ket{3}$}]        (3)       [right=of 2] {1};
\node[roundnode,label=below:{$\ket{4}$}]        (4)       [right=of 3] {1};
\node[roundnode,label=below:{$\ket{5}$}]		(5)       [right=of 4] {2};
\draw[ultra thick,->] (0.east) -- (1.west)  node[midway, below]{$\Delta_{01}$};
\draw[ultra thick,->] (2.east) -- (3.west)  node[midway, below]{$\Delta_{23}$};
\draw[ultra thick,->] (4.east) -- (5.west)  node[midway, below]{$\Delta_{45}$};
\draw[ultra thick,->] (0.north east) .. controls +(.6,.6) and +(-.6,.6) ..  (4.north west)  node[midway, above]{$\Delta_{04}$};
\end{tikzpicture}
\end{center} 
The next term to be incorporated, $V_{13}\ket{1}\!\bra{3}$, likewise requires merging of graph components, yielding (weights not shown from hereon):
\begin{center}
\begin{tikzpicture}[
roundnode/.style={circle, draw=black!60, fill=black!5, very thick, minimum size=8mm},
node distance=7mm,]
\node[roundnode,label=below:{$\ket{0}$}]        (0)                              {0};
\node[roundnode,label=below:{$\ket{1}$}]        (1)       [right=of 0] {1};
\node[roundnode,label=below:{$\ket{2}$}]		(2)       [right=of 1] {1};
\node[roundnode,label=below:{$\ket{3}$}]        (3)       [right=of 2] {2};
\node[roundnode,label=below:{$\ket{4}$}]        (4)       [right=of 3] {1};
\node[roundnode,label=below:{$\ket{5}$}]		(5)       [right=of 4] {2};
\draw[ultra thick,->] (0.east) -- (1.west);
\draw[ultra thick,->] (2.east) -- (3.west);
\draw[ultra thick,->] (4.east) -- (5.west);
\draw[ultra thick,->] (0.north east) .. controls +(.6,.6) and +(-.6,.6) ..  (4.north west);
\draw[ultra thick,->] (1.north east) .. controls +(.4,.4) and +(-.4,.4) ..  (3.north west);
\end{tikzpicture}
\end{center}
The remaining two drive terms would violate the integer constraint and hence cannot be included,  
\begin{center}
\begin{tikzpicture}[
roundnode/.style={circle, draw=black!60, fill=black!5, very thick, minimum size=8mm},
node distance=7mm,]
\node[roundnode,label={[label distance=.25cm]below:{$\ket{0}$}}]        (0)                              {0};
\node[roundnode,label={[label distance=.25cm]below:{$\ket{1}$}}]        (1)       [right=of 0] {1};
\node[roundnode,label={[label distance=.25cm]below:{$\ket{2}$}}]		(2)       [right=of 1] {1};
\node[roundnode,label={[label distance=.25cm]below:{$\ket{3}$}}]        (3)       [right=of 2] {2};
\node[roundnode,label={[label distance=.25cm]below:{$\ket{4}$}}]        (4)       [right=of 3] {1};
\node[roundnode,label={[label distance=.25cm]below:{$\ket{5}$}}]		(5)       [right=of 4] {2};
\draw[ultra thick,->] (0.east) -- (1.west);
\draw[ultra thick,->] (2.east) -- (3.west);
\draw[ultra thick,->] (4.east) -- (5.west);
\draw[ultra thick,->] (0.north east) .. controls +(.6,.6) and +(-.6,.6) ..  (4.north west);
\draw[ultra thick,->] (1.north east) .. controls +(.4,.4) and +(-.4,.4) ..  (3.north west);
\draw[ultra thick,dashed,->] (2.south east) .. controls +(.4,-.4) and +(-.4,-.4) ..  (4.south west);
\draw[ultra thick,dashed,->] (3.east) -- (4.west);
\end{tikzpicture}
\end{center}
Such dashed edges appear when graph cycles emerge that do not adhere to the requirement of zero-cyclicity \cite{Einwohner1976} which can be understood as follows. Consider a clockwise traversal of a graph cycle. Let $P$ denote the number of edges in the cycle where the final state has a higher index than the initial state, and $Q$ the corresponding number of edges where the final state has the lower index. If $P - Q \neq 0$, then a dashed edge cannot be avoided. This will always be the case for a cycle with an odd number of edges, such as for a three-level system.  

\section{Equivalence of Graph Merging Choices \label{appendix B}}

In this appendix, we show that the freedom in how to merge two graph components leads to equivalent graphs. We encountered an example of this in App. \ref{appendix A}, where merging of two disconnected graph components could either be achieved by up-shifting integers in one component, or down-shifting them in the other. We will show that both choices lead to equivalent effective Hamiltonians, differing only in an irrelevant global energy shift. 

Let us denote the integers associated with a graph component (set of vertices connected by edges) as a vector, $(k_{1}, k_{2}, k_{3}, \ldots)$,
so that $k_{n}$ is the integer chosen for the $n^{th}$ eigenstate (vertex) in the graph component. For the issue of merging, we now consider integers associated with two graph components, $\vec{a}$ and $\vec{b}$. Since the two graph components are separate before merging, the vectors $\vec{a}$ and $\vec{b}$ are spanned by disjoint sets of Cartesian basis vectors. In particular, if $\vec{a}\in \mathbb{Z}^{N}$ and $\vec{b}\in\mathbb{Z}^{M}$, then the merged graph's integers simply form a vector in $\mathbb{Z}^{N}\oplus\mathbb{Z}^{M}$. 

The freedom in merging consists of either up-shifting one component by some integer $k\in\mathbb{Z}$, or down-shifting the other by $-k$. Up-shifting component $\vec{a}$ by $k$ amounts to $\vec{a} \rightarrow \vec{a}' \equiv \vec{a} + \vec{k}_N$, with $\vec{k}_N = k(1, 1, \dots)\in\mathbb{Z}^N$. The merged graph vector representation is then 
\begin{equation*}
\vec{a}'\oplus\vec{b} = (
a_{1} + k, a_{2} + k, \ldots, b_{1}, b_{2}, \ldots
).
\end{equation*}
On the other hand, merging the graph components by down-shifting $\vec{b}$ yields $\vec{b} \rightarrow \vec{b}'\equiv \vec{b} - \vec{k}_M$ with the merged-graph representation
\begin{equation*}
\vec{a}\oplus\vec{b}' = (
a_{1}, a_{2}, \ldots, b_{1} - k, b_{2} - k, \ldots
).
\end{equation*}
Subtracting these two vectors gives 
\begin{equation*}
\vec{a}'\oplus\vec{b}-\vec{a}\oplus\vec{b}' = \vec{k}_{N+M} = k( 1, 1, \dots,  1, 1, \dots).
\end{equation*}
Therefore, the only difference between these two merge choices is a global shift of every state's integer by $k$. Accordingly, the two rotating-frame generators only differ by $\Omega \rightarrow \Omega + k\openone$, and the resulting effective Hamiltonians are the same up to an irrelevant global shift, $h \rightarrow h - k\,\omega_d\openone$. 
In conclusion, the two graph-merging choices lead to physically fully equivalent descriptions of the system.  

\section{Comments on Computational and Memory Efficiency\label{appendixC}} 
We briefly discuss the efficiency of the adaptive-RWA method and the Floquet-map method proposed by Hartmann et al.\ \cite{Hartmann2017}. The adaptive RWA is an iterative scheme, where each iteration involves solving a set of $D(D-1)/2$ inhomogeneous matrix equations $\eqref{eq:20}$ (corresponding to the given drive terms). For dense matrices, the computation time for solving Eq.\ $\eqref{eq:20}$ scales as $D^{6}$ using a direct method such as LU decomposition, so the total computation time $\tau$ scales as $\tau \sim N D^{8}$, where $N$ is the needed number of iterations. The superoperator in Eq.\ $\eqref{eq:20}$ is typically sparse, so the scaling can be improved using an iterative method such as least-squares minimization. Calculation of the Floquet map, i.e., the single-period dissipative propagator, involves time-evolving the $D(D+1)/2$ Hubbard operators $\ket{n}\!\bra{m}$ over one drive period $T$. The corresponding computation time additionally depends on the time-step size $\Delta t$ used by the ODE solver. The scaling in $D$ of each time-step depends on whether an implicit or explicit ODE solver is used ($D^6$ or $D^4$, respectively). This results in the scaling $\tau \sim (T/\Delta t) D^8$ [or $(T/\Delta t) D^6$]. The scaling of $\tau$ with $D$ will generally be somewhat more favorable for both methods, since the superoperators involved usually are not dense. While it is difficult to make general statements comparing the computational efficiency of the two methods, for the concrete example of the heavy fluxonium-resonator system we found the adaptive-RWA method to be more efficient than the Floquet-map method. 

Memory requirements also scale differently for the two methods. Adaptive RWA requires storage of the sparse superoperator $\mathbb{L}_{0} - i\,k_{nm}\omega_{d}\openone$ in Eq.\ $\eqref{eq:20}$.  The Floquet map method, on the other hand, requires storage of the single-period propagator, which is generally a dense $D^2\times D^2$ matrix, posing a possible memory bottleneck as Hilbert-space size increases.


\bibliographystyle{apsrev4-1}
\bibliography{refs}   

\end{document}